\documentstyle[preprint,aps]{revtex}

\begin{document}
\draft
\title{Broken rotation symmetry in the
fractional quantum Hall system}
\author{K. Musaelian and Robert Joynt}
\address{
Department of Physics\\
University of Wisconsin-Madison\\
1150 University Avenue\\
Madison, Wisconsin 53706\\}
\date{\today}
\maketitle
\begin{abstract}
We demonstrate that the two-dimensonal electron system in
a strong perpendicular magnetic field has
stable states which break rotational but
{\it not} translational symmetry.
The Laughlin fluid becomes unstable to these states
in quantum wells whose thickness exceeds
a critical value which
depends on the electron density.  The order parameter
at 1/3 reduced density resembles that of a nematic
liquid crystal, in that a residual two-fold rotation
axis is present in the low symmetry phase.
At filling factors 1/5 and 1/7, there are states
with four- and six-fold axes, as well.
We discuss the experimental detection of these phases.
\end{abstract}
\pacs{PACS Nos. 73.50.Jt, 71.45.-d, 72.15.Rn, 73.40.Kp}
\narrowtext

The construction of the Laughlin fluid and its identification
with the fractional quantum Hall effect (FQHE) marked the discovery of a
qualitatively new many-body phase. \cite{laughlin}
Until that discovery, it was generally assumed that the
only phases present in the two-dimensional electron system were
the usual liquid phase and the Wigner solid, both of which
are present in the classical one-component Coulomb plasma phase diagram.
A hexatic phase is also a possibility in this system \cite{morf}
as well as in the logarithmic potential case perhaps
more relevant to the FQHE. \cite{french} This theoretical background,
and the experimental discovery of phases
which seem to have properties unlike both
the Laughlin liquid and the Wigner solid, \cite{goldman} (''Hall insulator'')
led us to an investigation of the possibility of
broken rotational symmetry (BRS) in the two-dimensional
electron system in strong field.  We found that
several liquid crystal-like phases occur, though these appear to
resemble more closely nematic than hexatic phases.

We begin with the case where the electron density
$n$ satisfies $n=\frac{1}{3} \frac{1}{2 \pi \ell^2}, $
where $\ell$ is the magnetic length:
$\ell^{-2} = eB/\hbar c$ and B is the external field,
taken to be in the negative z-direction.  In this paper we work in the
limit of large field.
This is the density for the 1/3 quantum Hall state.
Consider the following wavefunction for the
disc-shaped system:
\begin{equation}
\Psi_{\alpha}({z_i})
=\prod_{i<j}^N [(z_i-z_j)(z_i-z_j-\alpha \ell)(z_i-z_j+\alpha \ell)]
\exp(-\sum_{i}|z_i|^2/4 \ell^2).
\label{eq:psi}
\end{equation}
Here $z_i = x_i + i y_i$ and i,j are particle indices.
$\alpha$ is a complex number.  This wavefunction
is antisymmetric in the particle indices for all $\alpha$,
lies entirely in the lowest Landau level,
and reduces to the Laughlin wavefunction at $\alpha = 0$.
It also shares with the Laughlin wavefunction the characteristic
of having uniform density (far from the edges of the disc).
However, the two-particle correlation function is a different matter:
\begin{equation}
g_{\alpha}(\vec{r}) = \frac{N(N-1)}{n^2}
                      \frac{(\prod_{i>2}^N \int d^2z_i) |\psi_{\alpha}|^2}
                           {(\prod_{i} \int d^2z_i) |\psi_{\alpha}|^2}.
\label{eq:corr}
\end{equation}
In this equation $\vec{r} = (x_1-x_2, y_1 - y_2) $,
and N is the total number of electrons .
Translational symmetry is not
broken for the small values of $\alpha$ with which we are concerned,
as we shall demonstrate below.  Thus  $g_{\alpha}$ is a
function only of the difference variable $\vec{r}$.
In contrast to the Laughlin liquid, however,
it does depend on the direction of $\vec{r}$.
Let us take $\alpha$ to be real.  Then the equivalent
classical plasma interaction corresponding to $\Psi_{\alpha}$
is a logarithmic interaction between rodlike charged objects
lying along the x-axis.  Accordingly, $\Psi_{\alpha}$
represents a BRS state. \cite{note}
In general $({\rm Re} \alpha, {\rm Im} \alpha)$ is a director,
not a vector, order parameter, since $\Psi_{\alpha} = \Psi_{-\alpha}$.
Thus $g_{\alpha}(\vec{r})$ does not have full rotation symmetry
for $\alpha \neq 0$ but does satisfy $g_{\alpha}(x,y)
=g_{\alpha}(-x,y)=g_{\alpha}(x,-y)=g_{\alpha}(-x,-y)$.

To give a physical picture of this state, we display a typical configuration
in a Monte Carlo simulation governed by the probability
distribution $|\Psi_{\alpha}|^2$ in Fig. \ref{fig:conf},
with $\alpha$ = 3.2.
One notes immediately the stripes along the director,
reminiscent of a nematic state.  This value of $\alpha$
is unphysically large and is chosen for
purposes of illustration only.

Under what conditions is such a BRS state stable?
Or, given an interelecron potential $V(\vec{r}_i-\vec{r}_j)$,
when do we have $ U_{\alpha} = \langle\Psi_{\alpha}|V|\Psi_{\alpha}\rangle <
U_0 = \langle\Psi_{0}|V|\Psi_{0}\rangle$ ?
It is easy to see, expanding the polynomial
in Eqn.\ \ref{eq:psi}, and taking the limit
$|\vec{r}|\rightarrow 0$ in Eqn. \ref{eq:corr}
that $g_{\alpha}(\vec{r}) \sim r^2$
rather than $g_{\alpha}(\vec{r}) \sim r^6$.
Thus for very short-range V, the Laughlin state is
always favored, as is well known. \cite{talapov} \cite{trugman}

In order to compare the $\alpha = 0$ with the $\alpha \neq 0$ states we
performed Monte Carlo simulations of the equivalent classical plasmas
with 200 particles in the disk geometry,
and compared the energies of candidate ground states.
(We believe that improvement of the wavefunction
by, for example, quantum Monte Carlo techniques \cite{ceperley}
will not affect energy {\it differences} very much.)
In order to minimize the finite size effect we compute the energies of only
25 particles closest to the center of the disk. We have checked that this
procedure gives the accepted value for the energy per particle in the case of a
pure Coulomb potential and $\alpha$ = 0.

{}From the computations we found that the Laughlin state is
favored for a pure Coulomb potential
$V(r)=e^2/\varepsilon r$. The actual potential between electrons
in a real two-dimensional layer of finite thickness
is softer, owing to the averaging of the charge
density over the third dimension.
This has been discussed in detail
by numerous authors and the chice of potential depends on
the shape of the well.
We shall take the simple form of Zhang and Das Sarma. \cite{zds}
They showed that $V(r)=e^2/\varepsilon (r^2 + \lambda^2)^{1/2}$ is a
reasonable approximation for a square quantum well and that $\lambda \approx
0.2 t$,
where t is the layer thickness of the well.
To understand the effect that this modification of the
potential will have, note that the total energy is given by
\begin{equation}
U_{\alpha} = \frac{ne^2}{2\varepsilon} \int \frac{d^2r}{(r^2+\lambda^2)^{1/2}}
             [g_{\alpha}(\vec{r})-1].
\end{equation}
We now plot
the angle averaged correlation functions for different values of
$\alpha$ in Fig.\ \ref{fig:gofr}.
There is incipient solid order as $\alpha$ increases,
in that there is much more tendency towards being
able to identify shells of neighbors.
At short distances, the correlation function is
proportional to $r^2$, just as that for the Wigner crystal is,
and in contrast to the $r^6$ behavior of the Laughlin state.
Overall, the the softer potential favors finite $\alpha$:
the correlation hole has a larger effective radius,
even though it is not as ''deep''. The difference between the energy per
particle for a BRS state with $\alpha = 1$ and the Laughlin state for different
values of $\lambda$ is plotted on Fig.\ \ref{diff:lambda}.
There is thus a critical value $\lambda_c$ at
which the system undergoes a transition to
finite $\alpha$.
We compute this to be
$\lambda_c = 4.1 \pm 1..5$, which corresponds to a thickness
of t = 1600\AA, when B = 10T.  This transition is second order,
unlike the transition to the Wigner solid,
which is probably first order.  It has recently been pointed out that
changing t can induce the liquid-solid transition by a
mechanism similar to that proposed here. \cite{sankar}
The critical value of t for the m=1/3 density is
similar to that computed here, suggesting that the
energy balance between Laughlin, BRS, and crystalline states
is a subtle one.  We expect that the
BRS state occupies a fairly narrow range of parameter
space between the liquid and the crystal states, by
analogy with hexatic phases.  It is clear, in any case, that this
range of thickness values is experimentally accessible. \cite{exp}
The energy balance between the BRS and Wigner crystal state is
currently under investigation.  \cite{future}

The correlations are oscillatory even to infinite
distances in a true crystalline state.  Inspection of
Fig.\ \ref{fig:gofr} shows that
the correlation function for the BRS state is
flat at large distances, demonstrating that
long-range translational symmetry is not
broken, and justifying the identication of these
states as liquid crystal states.

We may construct similar wavefunctions for m=5 and m=7.
\begin{equation}
\Psi_{\alpha}({z_i})
=\prod_{i<j}^N [(z_i-z_j)(z_i-z_j-\alpha \ell)(z_i-z_j+\alpha \ell)
(z_i-z_j - \beta \ell)(z_i-z_j+ \beta \ell)] \exp(-\sum_{i}|z_i|^2/4 \ell^2)
\label{eq:psi5}
\end{equation}
is an appropriate wavefunction for m=5 and
\begin{eqnarray}
\Psi_{\alpha}({z_i})
&=&\prod_{i<j}^N [(z_i-z_j)
(z_i-z_j-\alpha \ell)(z_i-z_j + \alpha \ell)
(z_i-z_j-\beta \ell)(z_i-z_j + \beta \ell) \\
&\times&
(z_i-z_j-\gamma \ell)(z_i-z_j + \gamma \ell)]
 \exp(-\sum_{i}|z_i|^2/4 \ell^2)
\label{eq:psi7}
\end{eqnarray}
is an appropriate wavefunction for m=7.
We have not yet investigated these wavefunctions for
all values of $\beta$ and $\gamma$.  Particularly interesting cases
are $\beta = i \alpha$ for m=5 and
$\beta = \omega \alpha, \gamma=\omega^2 \alpha$ for m=7
if $\omega$ is chosen as
$\exp(2\pi i/6)$.  The polynomial parts of the wavefunctions for this parameter
choice
may also be written as $\prod z(z^4-\alpha^4)$ (for m=5) and
$\prod z(z^6-\alpha^6)$ (for m=7), where $z=z_i-z_j$.
The correlation functions for the m=5 wavefunction have a four-fold rotation
axis and a six-fold rotation
axis for the m=7 wavefunction.  The incipient solid ordering again will
stabilize
these states as the well increases in thickness.  For
m=5 and m=7 we calculate critical values
$\lambda_c(m=5)=2.9 \pm 0.3$
and $\lambda_c(m=7)=2.1 \pm 1.7$, respectively.
The latter is a state which resembles the hexatic state
of two-dimensional fluids. The hexatic state, however, does
not pick out particular directions in space while the
m=7 state does. The m=5 state is somewhat similar to a biaxial nematic.

The director $\vec{n}$ = (Re $\alpha$, Im $\alpha$) is the order parameter
whose appearance signals the appearance of BRS. The
states characterized by $\vec{n}$ and -$\vec{n}$ are identical.
There is no independent inversion symmetry operation in our
two-dimensional system and thus the transition to the
low-symmetry phase is second-order, unlike the situation for
ordinary three-dimensional nematic systems.  Our Monte Carlo calculations
of energy as a function of $\alpha$ confirm this picture.
The Ginzburg-Landau energy is therefore
\begin{equation}
F=A(T,t)n^2 + B n^4 +  K_1 (\nabla \cdot \vec{n})^2
  +  K_2 (\nabla \times \vec{n})^2,
\end{equation}
where t is the thickness and T is the temperature.
We expect a second-order transition when T=0 and
our calculations have been carried out only at zero
temperature.  At finite temperature thermal fluctuations
are important.  Depending on the experimental situation
they may convert the transition to one
of the Kosterlitz-Thouless type.

The free emergy expression shows that twists of the order parameter are
possible and lead to textures but also to low energy excited states.
Nevertheless,
these excited states do not involve density changes and the state
as a whole is incompressible.  We conclude that the
FQHE still occurs in this gapless state.  The BRS state does not appear
to be a candidate for the Hall insulator phase.
The quasiparticle and quasihole excitations are still gapped
and their charges are the usual fractional ones.  Their
charge density profiles will have elliptical distortion.
Since the projected oscillator strength $f(\vec{k})$
and the projected static structure factor $S(\vec{k})$ depend
on the direction of $\vec{k}$, the magnetoroton excitations
with energy $E(\vec{k}) = f(\vec{k}) / S(\vec{k})$ have
dispersion which depends on the direction.

{}From the experimental point of view, it appears
that the chief difficulty in identifying the BRS
states lies in distinguishing them from the Laughlin state.
Correlation functions are anisotropic, but scattering experiments to test
this are difficult to perform in two-dimensional systems.
Tensor quantities such as the
conductivity have a characteristic anisotropy (birefringence):
$\sigma_{xx}(\omega) \neq \sigma_{yy}(\omega)$,
except at zero frequency, when
$\sigma_{xx}(\omega=0) = \sigma_{yy}(\omega=0) = 0$,
as usual.
Propagation of surface acoustic waves or measurements of the
microwave conductivity, perhaps with the simultaneous application of a current
to eliminate domain effects,
may be tools which can probe such an anisotropy,
and test for the existence of the BRS states.

We wish to thank A.V. Chubukov and M.B. Webb for useful discussions.
This work was supported by the National Science Foundation under Grant No.
DMR-9214739.

\begin{figure}
\caption{Typical configuration of the particles in the Monte Carlo simulation
of the state given by Eqn.\ (1) with $\alpha = 3.2$  The anisotropy of the
correlations is clearly evident.}
\label{fig:conf}
\end{figure}

\begin{figure}
\caption{Angle averaged pair correlation function for the m=1/3 state for
$\alpha = 0$ and $\alpha = 3.2$.  The $\alpha = 3.2$ state shows
incipient crystalline behavior at short distances and
liquid-like behavior at long distances.}
\label{fig:gofr}
\end{figure}

\begin{figure}
\caption{The difference between the energies per particle in a BRS state with
$\alpha = 1$ and the Laughlin state as a function of $\lambda$,
which is a measure of the well thickness.}
\label{diff:lambda}
\end{figure}


\begin{references}
\bibitem{note} A more general approach to the problem would be to
replace $\alpha$ by $\alpha_{ij}$ and let the $\alpha_{ij}$ be drawn from
an appropriate probability distribution.  This would allow
systematic improvement of the Laughlin function for a given potential.
The BRS state would then appear as a real normal-nematic transition in the
equivalent
plasma problem.
\end{references}
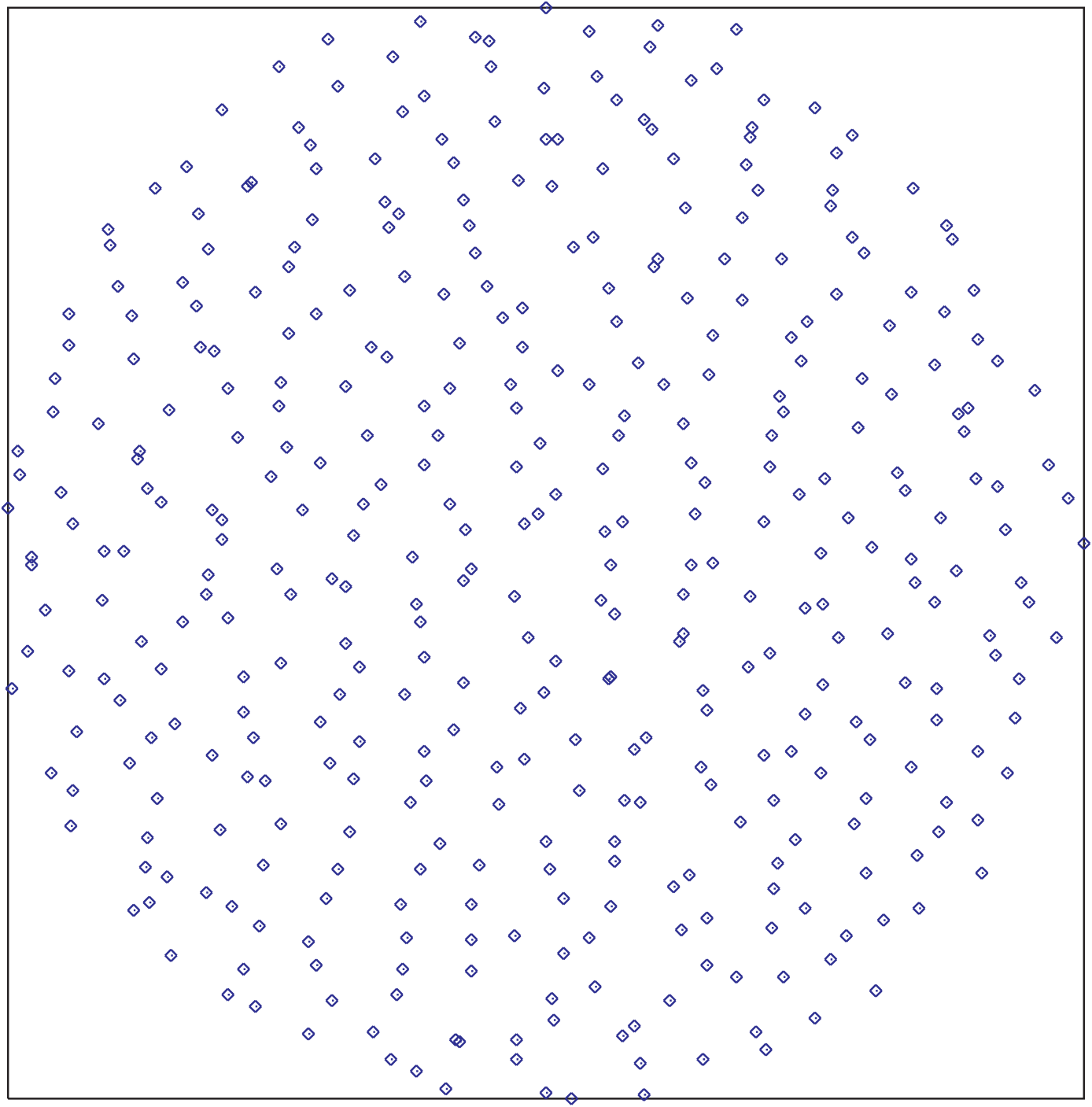
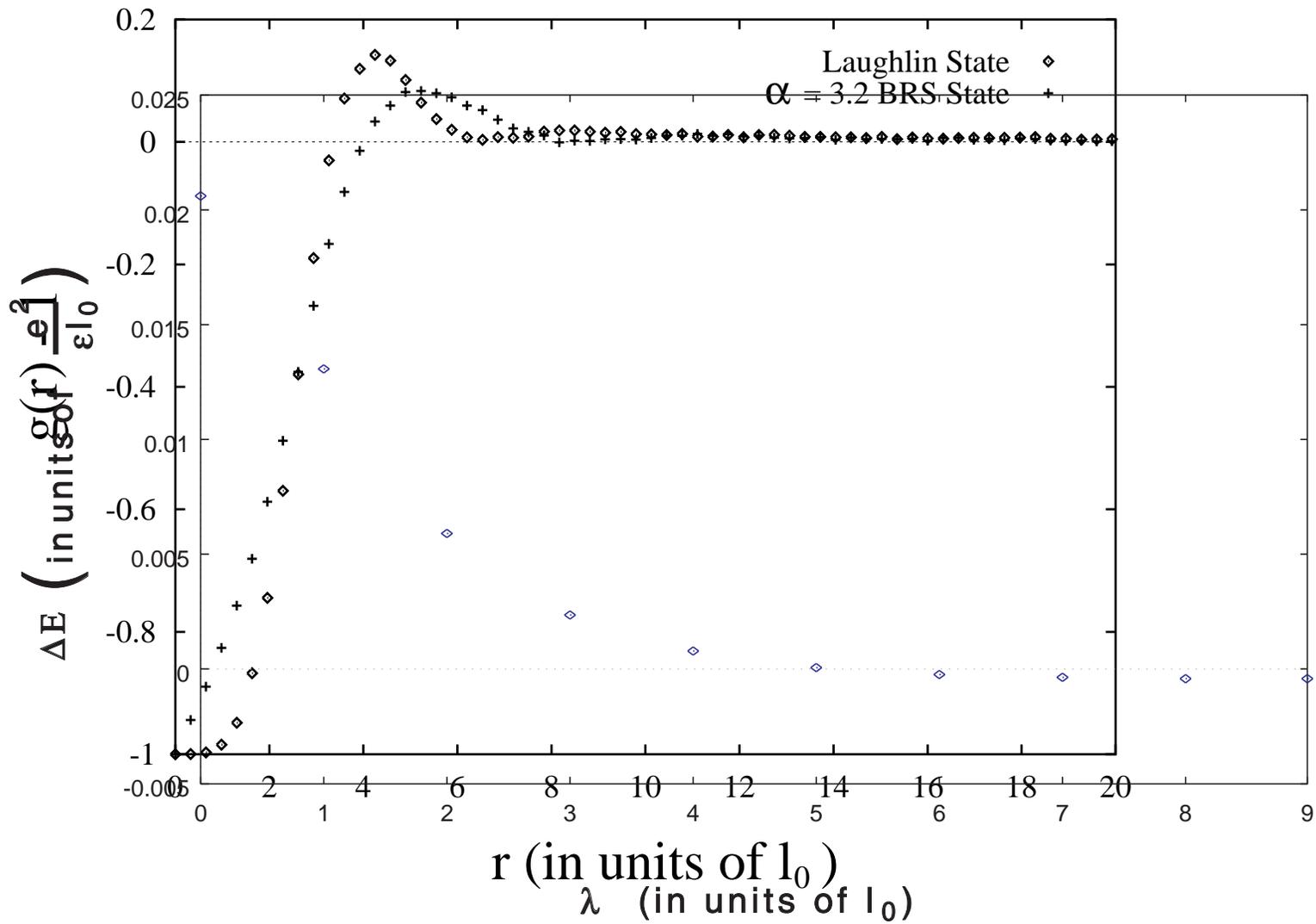
\end{document}